\documentclass[final,3p,times]{elsarticle}
\usepackage{amssymb}
\usepackage{amsmath}
\usepackage{lineno,hyperref}

\modulolinenumbers[5]
\biboptions{sort&compress}
\journal{Annals of Physics}
\hypersetup{colorlinks=true,linkcolor=blue,citecolor=blue,filecolor=blue,urlcolor=blue,breaklinks=true}

\begin{document}

\title{New solutions of the $D$-dimensional Klein-Gordon equation via
mapping onto the nonrelativistic one-dimensional Morse potential}
\author[1,2]{M. G. Garcia}
\ead{marcelogarcia82@gmail.com}
\author[3,4]{A. S. de Castro}
\ead{castro@pq.cnpq.br}
\author[5]{L. B. Castro}
\ead{luis.castro@pq.cnpq.br}
\author[4]{P. Alberto}
\ead{pedro.alberto@uc.pt}

\address[1]{UNICAMP, Universidade de Campinas, Departamento de Matem\'{a}tica Aplicada,
	13081-970, Campinas, SP, Brazil}

\address[2]{ITA, Intituto Tecnol\'{o}gico de Aeron\'{a}utica, Departamento de F\'{\i}sica,
	12228-900, S\~{a}o Jos\'{e} dos Campos, SP, Brazil}

\address[3]{UNESP, Universidade Estadual Paulista, Campus de Guaratinguet\'{a}, Departamento de F\'{\i}sica e
	Qu\'{\i}mica, 12516-410, Guaratinguet\'{a}, SP,  Brazil}

\address[4]{CFisUC, University of Coimbra, Physics Department,
	P-3004-516, Coimbra, Portugal}

\address[5]{UFMA, Universidade Federal do Maranh\~{a}o,
	Campus Universit\'{a}rio do Bacanga, Departamento de F\'{\i}sica, 65080-805, S\~{a}o Lu\'{\i}s, MA,
	Brazil}

\begin{abstract}
New exact analytical bound-state solutions of the $D$-dimensional
Klein-Gordon equation for a large set of couplings and potential functions
are obtained via mapping onto the nonrelativistic bound-state solutions of
the one-dimensional generalized Morse potential. The eigenfunctions are
expressed in terms of generalized Laguerre polynomials, and the
eigenenergies are expressed in terms of solutions of irrational equations at
the worst. Several analytical results found in the literature, including the
so-called Klein-Gordon oscillator, are obtained as particular cases of this
unified approach.
\end{abstract}

\maketitle

\section{Introduction}
The generalized Morse potential $Ae^{-\alpha x}+Be^{-2\alpha x}$ \cite{Tezcan2009,c9a1c724ae714580b3931e1f2509b0bb, BagrovGitman199002,chetouani1992algebraic,chetouani1993construction, arda2012exact}, the singular harmonic oscillator (SHO) $Ax^{2}+Bx^{-2}$ \cite{BagrovGitman199002}, \cite{landau1965quantum,gol2012problems,constantinescu2013problems,Haar201408,ballhausen1988note,ballhausen1988step,palma2003one,palma2003addendum,dong2005exact,QUA:QUA21103,singh2006ladder,Dong201011,dong2007exact,patil2007uncertainty,ikhdair2009exactly,pimentel2013singular} and the singular Coulomb potential (SCP)
\thinspace $Ax^{-1}+Bx^{-2}$ \cite{BagrovGitman199002}, \cite{landau1965quantum,gol2012problems,constantinescu2013problems,Haar201408}, \cite{dong2007exact}, \cite{ikhdair2009exactly}, 
\cite{Condon1929, hall1998smooth, oyewumi2005analytical,ikhdair2008exact,ho2009simple, Agboola2011,pimentel2014brief,das2015exact} have played an important role
in atomic, molecular and solid-state physics. Bound states for systems
modelled by those potentials are computed exactly in nonrelativistic quantum
mechanics. In a recent paper \cite{Nogueira2016}, it was shown that the Schr\"{o}%
dinger equation for all those exactly solvable problems mentioned above can
be reduced to the confluent hypergeometric equation in such a way that it
can be solved via Laplace transform method with closed-form eigenfunctions
expressed in terms of generalized Laguerre polynomials. Connections between
the Morse and those other potentials have also been reported. The Langer
transformation \cite{Langer1937} a nonunitary transformation consisting of a
change of function plus a map from the full line of a one-dimensional
problem to the half line appropriate to a radial $D$-dimensional problem, has been
a key ingredient for those connections. The three-dimensional Coulomb
potential has been mapped into the one-dimensional Morse potential and into
the three-dimensional singular Coulomb potential \cite{Langer1937}. The Morse
potential with particular parameters has been mapped into the
two-dimensional harmonic oscillator \cite{montemayor1983ladder} and into the
three-dimensional Coulomb potential \cite{lee1985hydrogen}. Later, the generalized Morse
potential was mapped into the three-dimensional harmonic oscillator and
Coulomb potentials \cite{doi:10.1119/1.14794,sun2008eigenenergies}. Furthermore, a certain mapping
between the Morse potential with particular parameters and a particular case
of the three-dimensional SCP has been found with fulcrum on the algebra $%
so\left( 2,1\right) $ and its representations \cite{cooper1994relation}. Recently, the introduction of two additional parameters into the Langer transformation provided opportunity to bound-state solutions of the SHO and SCP in arbitrary
dimensions be generated in a simple way from the bound states of the
one-dimensional generalized Morse potential with straightforward
determination of the critical attractive singular potential and the proper
boundary condition on the radial eigenfunction at the origin \cite{Nogueira2016a}.
More general connections for an extended Schr\"{o}dinger equation in a
position-dependent mass background have involved the three-dimensional
nonsingular Coulomb potential \cite{bagchi2006morse}, and the SHO and SCP in arbitrary
dimensions \cite{doi:10.1063/1.2838314}.

Relativistic effects on systems described by SHO and SCP can be approached
in the context of the Klein-Gordon (KG) theory. Actually, particular cases
of SHO and SCP under the umbrella of the KG theory have been studied in the
literature for different couplings, space dimensions and methodologies. The free KG equation is expressed as $\left( p^{\mu }p_{\mu
}-m^{2}c^{2}\right) \Psi =0$ with the linear momentum operator given by $%
p^{\mu }=i\hbar \partial ^{\mu }$. Vector and scalar interactions are
considered by replacing $p^{\mu }$ by $p^{\mu }-V^{\mu }/c$ and $m$ by $%
m+V_{s}/c^{2}$, respectively. From now on the time component of the vector
potential will be denominated by $V_{v}$ in this paper. The equal-magnitude
mixing of vector and scalar couplings for arbitrary angular momentum has
been considered for the harmonic oscillator \cite{1009-1963-11-8-301,1009-1963-13-3-003,zhang2009dynamical}, for $S$%
-waves of the SHO \cite{doi:10.1063/1.4902270} and for arbitrary angular momentum for the
pseudo-harmonic potential (a special particular case of the SHO) \cite{ikhdair2012effects}%
. The Coulomb potential has been studied as a vector coupling \cite{nieto1979hydrogen,fleischer1984bound,greiner1990relativistic}, scalar coupling \cite{fleischer1984bound,greiner1990relativistic}, mixing of vector and
scalar couplings with equal magnitudes \cite{zhang2009dynamical}, and an arbitrary mixing
of vector and scalar couplings \cite{dong2011wave,Hassanabadi2011,saad2008klein}. $S$-waves of the SCP
with arbitrary mixing of vector and scalar potentials have been considered 
\cite{kocak2007bound}, and for arbitrary angular momentum with equal magnitudes of
those potentials in the case of the Kratzer potential (a particular case of
the SCP) \cite{berkdemir2007relativistic}. In Refs. \cite{saad2008klein} and \cite{wen2003bound}, the SCP for
arbitrary angular momentum with a mixing of vector and scalar couplings with
equal magnitudes has also been considered. Dubbed the KG oscillator, an
unconventional form of interaction is achieved by replacing $\overrightarrow{%
p}^{2}$ by $\left( \overrightarrow{p}+im\omega r\widehat{r}\right) \cdot
\left( \overrightarrow{p}-im\omega r\widehat{r}\right) $ in such a way that
one obtains the Schr\"{o}dinger equation with the harmonic oscillator
potential in a nonrelativistic scheme \cite{bruce1993klein}. This alternative form of
interaction spurred a big and still growing literature \cite{dvoeglazov1994comment,mirza2004klein,rao2007energy,jian2008klein,li2010wigner,chargui2010exact,chargui2011path,cheng2011mathcal,mirza2011relativistic,boumali2011comment,xiao2011klein,liang2012three,hassanabadi2014statistical,boumali2014klein,boumali2015one,bakke2015klein,vitoria2016relativistic,Vitoria2016}. It is interesting to remark that this bears a striking resemblance to the
Dirac oscillator \cite{kukulin1991dirac}, where a harmonic oscillator potential is
introduced via a radial linear tensor coupling in the Dirac equation. More
recent literature treats the KG oscillator in two space dimensions under the
influence of a scalar Coulomb potential \cite{bakke2015klein}, a scalar linear
potential \cite{vitoria2016relativistic}, and a mixing of scalar linear and vector Coulomb
potentials \cite{Vitoria2016}, with solutions allowed only for certain parameters
of the KG oscillator and eigenfunctions expressed in terms of Heun
biconfluent functions. One caveat: in Ref. \cite{chargui2011path}, the authors
misidentified the one-dimensional KG equation with the space component of
the linear vector potential (minimally coupled) as the KG oscillator.

In the present paper, the method developed in \cite{Nogueira2016a} is extended to a
modified $D$-dimensional KG equation featuring a vector interaction
nonminimally coupled. Actually, this kind of generalized KG equation has the
same form as that one for the physical component of the five-component
spinor in a one-dimensional space \cite{cardoso2010nonminimal} and also in a three-dimensional
space \cite{castro2011spinless} appearing in the scalar sector of the Duffin-Kemmer-Petiau
theory. Specially noteworthy is the inclusion of the KG oscillator in that
framework. This extension of the method used in Ref. \cite{Nogueira2016a} is an
interesting way of providing a unified treatment of many known relativistic
problems via a mapping onto a unique well-known one-dimensional
nonrelativistic problem, allowing to obtain some new exact analytical
bound-state solutions for a large class of problems including new types of
couplings and potential functions. We highlight vector-scalar SHO plus
nonminimal vector Cornell potentials and nonminimal vector Coulomb (space
component) and harmonic oscillator (time component) potentials,
vector-scalar Coulomb plus nonminimal vector Cornell potentials and
nonminimal vector shifted Coulomb potentials, vector-scalar SCP plus
nonminimal vector Coulomb potentials, and also the curious case of a pure
nonminimal vector constant potential. Furthermore, we show that several
exactly soluble bound states explored in the literature are obtained as
particular cases of those cases. In all those circumstances the
eigenfunctions are expressed in terms of the generalized Laguerre
polynomials and the eigenenergies are expressed in terms of irrational
equations.

The paper is organized as follows. In Sec. 2 we review, as a background, the
generalized Morse potential in the Schr\"{o}dinger equation. The $D$%
-dimensional KG equation with vector, scalar and nonminimal vector\
couplings, its connection with the generalized Morse potential and the
proper form for the potential functions, are presented in Sec. 3. In Sec. 4
we present some concluding remarks.

\section{Nonrelativistic bound states in a one-dimensional generalized Morse
potential}

The time-independent Schr\"{o}dinger equation is an eigenvalue equation for
the characteristic pair $(E,\psi )$ with $E\in 
\mathbb{R}
$. For a particle of mass $M$ embedded in the generalized Morse potential it
reads 
\begin{equation}
\frac{d^{2}\psi \left( x\right) }{dx^{2}}+\frac{2M}{\hbar ^{2}}\left(
E-V_{1}e^{-\alpha x}-V_{2}e^{-2\alpha x}\right) \psi \left( x\right) =0,
\label{sch}
\end{equation}%
where $\alpha >0$. Bound-state solutions demand $\int_{-\infty }^{+\infty
}dx\,|\psi |^{2}=1$ and exist only when the generalized Morse potential has
a well structure ($V_{1}<0$ and $V_{2}>0$). The eigenenergies are then given
by (see, e.g., \cite{Nogueira2016}, \cite{Nogueira2016a})%
\begin{equation}
E_{n}=-\frac{V_{1}^{2}}{4V_{2}}\left[ 1-\frac{\hbar \alpha \sqrt{2MV_{2}}}{%
M|V_{1}|}\left( n+\frac{1}{2}\right) \right] ^{2},  \label{ENE}
\end{equation}%
with%
\begin{equation}
n=0,1,2,\ldots <\frac{M|V_{1}|}{\hbar \alpha \sqrt{2MV_{2}}}-\frac{1}{2}.
\label{cond}
\end{equation}%
This restriction on $n$ limits the number of allowed states and requires $%
M|V_{1}|/\left( \hbar \alpha \sqrt{2MV_{2}}\right) >1/2$ to make possible
the existence of bound states. On the other hand, with the substitutions%
\begin{equation}
\hbar \alpha s_{n}=\sqrt{-2ME_{n}},\quad \hbar \alpha \xi =2\sqrt{2MV_{2}}%
\,e^{-\alpha x},  \label{xi}
\end{equation}%
the eigenfunctions are expressed in terms of the generalized Laguerre
polynomials as%
\begin{equation}
\psi _{n}\left( \xi \right) =N_{n}\,\xi ^{s_{n}}e^{-\xi /2}L_{n}^{\left(
2s_{n}\right) }\left( \xi \right) ,  \label{psi}
\end{equation}%
where $N_{n}$ are arbitrary constants.

\section{The $D$-dimensional KG equation}

Incorporating the KG oscillator as a particular case, a generalized
Lorentz-covariant KG equation for a particle of mass $m$ is written (with $%
\hbar =c=1$) as

\begin{equation}
\left[ \left( p^{\mu }-V^{\mu }-iU^{\mu }\right) \left( p_{\mu }-V_{\mu
}+iU_{\mu }\right) -\left( m+V_{s}\right) ^{2}\right] \Psi =0.  \label{DKP}
\end{equation}%
In the absence of a scalar coupling, Eq. (\ref{DKP}) has the same form as that one for the physical component of
the five-component spinor in a one-dimensional space \cite{cardoso2010nonminimal} and also in a three-dimensional
space \cite{castro2011spinless} appearing in the scalar sector of the
Duffin-Kemmer-Petiau theory. A continuity equation of the form $\partial
_{\mu }J^{\mu }=0$ is satisfied with the current density $J^{\mu }$
proportional to $i\Psi ^{\ast }\overleftrightarrow{\partial ^{\mu }}\Psi
-2V^{\mu }|\Psi |^{2}$. In contrast to $V^{\mu }$, the vector potential $%
U^{\mu }$ is not minimally coupled. Furthermore, invariance under the
time-reversal transformation demands that $V^{\mu }$ and $U^{\mu }$ have
opposite behaviours. Similarly to the scalar potential, the nonminimal
vector potential does not couple to the charge because it does not change
its sign under the charge-conjugation operation ($\Psi \rightarrow \Psi
^{\ast },V^{\mu }\rightarrow -V^{\mu }$). In other words, $U^{\mu }$ and $%
V_{s}$ do not distinguish particles from antiparticles and so the system
does not exhibit Klein's paradox in the absence of the minimal coupling.
Note also that $J^{\mu }\rightarrow -J^{\mu }$ under charge conjugation, as
should be expected for a charge current density.

At this stage, we make $\vec{V}=\vec{0}$ (the space component of the minimal
coupling can be gauged away for spherically symmetric potentials) and
consider only time-independent potentials in such that the factorization $%
\Psi \left( \vec{r},t\right) =e^{-i\varepsilon t}\Phi \left( \vec{r}\right) $
yields%
\begin{equation}
\left\{ \left( \vec{p}-i\vec{U}\right) \cdot \left( \vec{p}+i\vec{U}\right)
-U_{0}^{2}-\left[ \left( \varepsilon -V_{v}\right) ^{2}-\left(
m+V_{s}\right) ^{2}\right] \right\} \Phi =0,
\end{equation}%
or, equivalently, 
\begin{equation}
\left[ \nabla ^{2}-\left( \vec{\nabla}\cdot \vec{U}\right) -\vec{U}%
^{2}+U_{0}^{2}+\left( \varepsilon -V_{v}\right) ^{2}-\left( m+V_{s}\right)
^{2}\right] \Phi =0.  \label{eq1}
\end{equation}%
Eq. (\ref{eq1}) is an eigenvalue equation for the characteristic pair ($%
\varepsilon ,\Phi $) with $\varepsilon \in 
\mathbb{R}
$. From this equation, it is clear that the spectrum is distributed
symmetrically about $\varepsilon =0$ in the absence of the minimal coupling
(charge conjugation changes $\varepsilon $ by $-\varepsilon $). Because the
charge density is proportional to $(\varepsilon -V_{v})|\Phi |^{2}$, bound
states demand $\Phi \rightarrow 0$ as $r\rightarrow \infty $. In spherical
coordinates of a $D$-dimensional space, the position vector is $\vec{r}%
=\left( r,\Omega \right) $, where $\Omega $ denotes a set of $D-1$ angular
variables.

For spherically symmetric potentials, $V_{v}\left( \vec{r}\right)
=V_{v}\left( r\right) $, $V_{s}\left( \vec{r}\right) =V_{s}\left( r\right) $%
, $U_{0}\left( \vec{r}\right) =V_{0}\left( r\right) $ and $\vec{U}\left( 
\vec{r}\right) =V_{r}\left( r\right) \hat{r}$, one can write%
\begin{equation}
\Phi \left( \vec{r}\right) =\,\frac{u(r)}{r^{k}}Y\left( \Omega \right) ,
\label{uy}
\end{equation}%
where $Y$ denotes the hyperspherical harmonics labelled by $2k$ quantum
numbers (see, e.g. \cite{Dong201011}, \cite{avery2012hyperspherical}),%
\begin{equation}
k=\left( D-1\right) /2,
\end{equation}%
and $u$ obeys the Schr\"{o}dinger-like radial equation%
\begin{equation}
\frac{d^{2}u\left( r\right) }{dr^{2}}+2M\left[ \widetilde{\varepsilon }%
-V\left( r\right) -\frac{L\left( L+1\right) }{2Mr^{2}}\right] u\left(
r\right) =0,  \label{Eq}
\end{equation}%
with $M$ denoting a positive parameter having dimension of mass. The
effective energy $\widetilde{\varepsilon }$ and the effective potential $V$
are expressed by 
\begin{eqnarray}
2M\widetilde{\varepsilon } &=&\varepsilon ^{2}-m^{2}  \notag
\label{def_eps_V} \\
2MV &=&V_{s}^{2}-V_{v}^{2}+2\left( mV_{s}+\varepsilon V_{v}\right) +\frac{%
dV_{r}}{dr}+2k\frac{V_{r}}{r}+V_{r}^{2}-V_{0}^{2},
\end{eqnarray}%
and $L$ in the centrifugal barrier potential takes the values%
\begin{equation}
L=l+k-1\quad \text{or}\quad L=-l-k,  \label{EQ2}
\end{equation}%
in which $l=0,1,2\ldots $ is the orbital momentum quantum number, and $%
u\rightarrow 0$ as $r\rightarrow \infty $ for bound-state solutions.

Following Ref. \cite{Nogueira2016a}, with effective potentials expressed by%
\begin{equation}
V\left( r\right) =Ar^{\delta }+\frac{B}{r^{2}}+C,\quad \delta =+2\text{ or }0%
\text{ or }-1  \label{vef}
\end{equation}%
the Langer transformation \cite{Langer1937} 
\begin{equation}
u(r)=\sqrt{r/r_{0}}\,\phi (x)\ ,\quad r/r_{0}=e^{-\Lambda \alpha x}\ ,
\label{Langer}
\end{equation}%
with $r_{0}>0$ and $\Lambda >0$, transmutes the radial equation (\ref{Eq})
into%
\begin{equation}
\frac{d^{2}\phi \left( x\right) }{dx^{2}}+2M\left\{ -\frac{\left( \Lambda
\alpha S\right) ^{2}}{2M}-\left( \Lambda \alpha r_{0}\right) ^{2}\left[
Ar_{0}^{\delta }e^{-\Lambda \alpha \left( \delta +2\right) x}+\left( C-%
\widetilde{\varepsilon }\right) e^{-2\Lambda \alpha x}\right] \right\} \phi
\left( x\right) =0,  \label{sch2}
\end{equation}%
with%
\begin{equation}
S=\sqrt{\left( l+k-1/2\right) ^{2}+2MB}.  \label{etil}
\end{equation}%
At this point, it is instructive to note that not only $L\left( L+1\right) $
is insensible to the different choices of $L$ as prescribed by (\ref{EQ2})
but also $S$. One can see from (\ref{sch2}) that there is no bound-state
solution if $V$ is an inversely quadratic potential ($A=0$, or $A\neq 0$
and $\delta =0$ or $\delta =-2$). A connection with the bound states of the
generalized Morse potential is reached only if the pair $(\delta ,\Lambda )$
is equal to $(2,1/2)$ or $(-1,1)$, and, as can be seen from the
identification of the first term in the curly brackets with the
(negative) energy parameter $E$ in (\ref{sch}) one has to have $S^{2}>0$.
Thus, 
\begin{equation}
2MB>-(2k-1)^{2}/4.  \label{beta}
\end{equation}%
Furthermore, from eq.~(\ref{sch2}), the asymptotic behaviour of $\phi (x)$
is $\phi \left( x\right) \underset{x\rightarrow +\infty }{\rightarrow }%
e^{-\Lambda \alpha Sx}$ such that, from (\ref{Langer})%
\begin{equation}
u\left( r\right) \underset{r\rightarrow 0}{\rightarrow }r^{1/2+S}.
\label{ur}
\end{equation}

Effective potentials of the general form (\ref{vef}) can be realized by the
following particular choices for the radial potentials in the Klein-Gordon
equation (\ref{eq1})%
\begin{eqnarray}
V_{r} &=&\beta _{r}/r+\gamma _{r}r^{\delta _{r}},\quad \delta _{r}=0\text{
or }1  \notag \\
V_{0} &=&\beta _{0}/r+\gamma _{0}r^{\delta _{0}},\quad \delta _{0}=0\,,1%
\text{ or }-1/2  \notag \\
V_{s} &=&\alpha _{s}/r^{2}+\beta _{s}/r+\gamma _{s}r^{2}, \\
V_{v} &=&\alpha _{v}/r^{2}+\beta _{v}/r+\gamma _{v}r^{2},  \notag
\end{eqnarray}%
with $\alpha _{v}^{2}=\alpha _{s}^{2}$ and $\alpha _{v}\beta _{v}=\alpha
_{s}\beta _{s}$ to eliminate inversely quartic and cubic terms in $V$,
respectively. This means $\alpha _{v}=\alpha _{s}=0$, or $\alpha _{v}=\pm
\alpha _{s}\neq 0$ with $\beta _{v}=\pm \beta _{s}$. On the other hand,
because in (\ref{vef}) $\delta$ can take just one value, when $%
\gamma_{s,v}\neq 0$, $\beta_{s,v}=0$ and vice-versa. For $\delta_0=-1/2$ one
has $\gamma_v=\gamma_s=\delta_r=\beta_0=0$.

The eigenvalue equation for the energy $\varepsilon$ is still obtained from
eq. (\ref{ENE}), but with different roles for $E$, $V_1$ and $V_2$ in eq. (%
\ref{sch}). Indeed, in the present case all these parameters can depend on
the energy $\varepsilon$ in two essential ways: either explicitly through
the term $\left( \Lambda \alpha r_{0}\right) ^{2} (C-\widetilde{\varepsilon }%
)$ in (\ref{sch2}), which can be equal to $V_1$ or to $V_2$, depending on
whether the pair $(\delta ,\Lambda )$ is equal to $(2,1/2)$ or $(-1,1)$,
respectively, or through the parameters $A$ and $B$ of the potential $V(r)$ (%
\ref{vef}) whenever the potential $V_v$ in (\ref{def_eps_V}) has the
correspondent radial dependence.

\subsection{The effective singular harmonic oscillator}

With $(\delta ,\Lambda )=(2,1/2)$ plus the definition $A=M\omega ^{2}/2$,
the identification of the bound-state solutions of Eq. (\ref{Eq}) with those
ones from the generalized Morse potential is done by choosing $V_{1}=-\alpha
^{2}r_{0}^{2}\left( \widetilde{\varepsilon }-C\right) /4$ and $V_{2}=\alpha
^{2}r_{0}^{4}M\omega ^{2}/8$, necessarily with $\widetilde{\varepsilon }>C$
and $\omega ^{2}>0$. With $\omega >0$ one can write, from eq.~(\ref{xi}),%
\begin{equation}
\xi =M\omega r^{2}.
\end{equation}%
Furthermore, (\ref{cond}) implies $\widetilde{\varepsilon }>C+\omega \left(
2n+1\right) $. Using (\ref{ENE}) and (\ref{etil}) one can write the complete
solution of the problem as%
\begin{eqnarray}  \label{tilde_eps}
\widetilde{\varepsilon } &=&C+\omega \left( 2n+1+S\right) \\
u(r) &=&Nr^{1/2+S}e^{-M\omega r^{2}/2}L_{n}^{\left( S\right) }\left( M\omega
r^{2}\right) ,
\end{eqnarray}%
where $\widetilde{\varepsilon }$ and $\omega $ depend on $\varepsilon $
because of eqs.~(\ref{def_eps_V}) and (\ref{vef}). The condition (\ref{cond}%
) means that 
\begin{equation}  \label{condSHO}
n\leq N=\bigg[\frac{M|V_{1}|}{\alpha \sqrt{2MV_{2}}}-\frac{1}{2}\bigg]= %
\bigg[\frac{\widetilde{\varepsilon }-C-\omega}{2\omega}\bigg]
\end{equation}
where $[x]$ stands for the largest integer less or equal to $x$. This means
that there is no limitation on the value of $N$, because it can be as large
as the energy can, which in turns means that $n$ in (\ref{tilde_eps}) has no
upper bound.

\subsubsection{Vector-scalar SHO plus nonminimal vector Cornell potentials}

An example of this class of solutions can be reached by choosing%
\begin{equation}
V_{r}=\frac{\beta _{r}}{r}+\gamma _{r}r,\quad V_{0}=\frac{\beta _{0}}{r}%
+\gamma _{0}r,\quad V_{s}=\frac{\alpha _{s}}{r^{2}}+\gamma _{s}r^{2},\quad
V_{v}=\pm V_{s}.  \label{P1}
\end{equation}%
This class represents a generalization of the cases found in \cite{1009-1963-11-8-301} ($%
V_{r}=V_{0}=0$ and $V_{v}=V_{s}=\gamma _{s}r^{2}$ in three dimensions), \cite%
{1009-1963-13-3-003} and \cite{zhang2009dynamical} ($V_{r}=V_{0}=0$ and $V_{v}=V_{s}=\gamma _{s}r^{2}$ in
two dimensions), \cite{doi:10.1063/1.4902270} ($V_{r}=V_{0}=0$ and $V_{v}=V_{s}=\alpha
_{s}/r^{2}+\gamma _{s}r^{2}$ for $S$-waves in three dimensions), \cite{ikhdair2012effects} (%
$V_{r}=V_{0}=0$ and the pseudo-harmonic potential in two dimensions) and 
\cite{bruce1993klein} ($V_{0}=V_{v}=V_{s}=0$ and $\beta _{r}=0$, as a matter of fact
the space component of a nonminimal vector anisotropic linear potential in
three dimensions).

The complete identification with the generalized Morse potential is done
with the identifications%
\begin{eqnarray}
M\omega  &=&\sqrt{\gamma _{r}^{2}-\gamma _{0}^{2}\pm 2\gamma _{s}\left(
\varepsilon \pm m\right) }  \notag \\
2MB &=&\beta _{r}\left( \beta _{r}-1+2k\right) -\beta _{0}^{2}\pm 2\alpha
_{s}\left( \varepsilon \pm m\right)  \\
2MC &=&\gamma _{r}\left( 2\beta _{r}+1+2k\right) -2\gamma _{0}\beta _{0}, 
\notag
\end{eqnarray}%
which lead, in general, using eq. (\ref{tilde_eps}), to an irrational
equation in $\varepsilon $%
\begin{equation}
\left( \varepsilon +m\right) \left( \varepsilon -m\right) -\gamma _{r}\left(
2\beta _{r}+1+2k\right) +2\gamma _{0}\beta _{0}=2\left( 2n+1+S\right) \sqrt{%
\gamma _{r}^{2}-\gamma _{0}^{2}\pm 2\gamma _{s}\left( \varepsilon \pm
m\right) },
\end{equation}%
or, more explicitly, using eq. (\ref{etil}), 
\begin{flalign}
\left( \varepsilon +m\right) &\left( \varepsilon -m\right) -\gamma _{r}\left(
2\beta _{r}+1+2k \right) +2\gamma _{0}\beta _{0}  \nonumber\\
  =&\,2\left( 2n+1+\sqrt{\left( l+k -1/2\right) ^{2}+
  \beta _{r}\left( \beta _{r}-1+2k \right) -\beta _{0}^{2}\pm
2\alpha _{s}\left( \varepsilon \pm m\right)}\right)\nonumber\\
\label{Na}&\times\sqrt{\gamma _{r}^{2}-\gamma _{0}^{2}\pm 2\gamma _{s}\left( \varepsilon \pm
m\right) }.
\end{flalign}

One notices immediately that, if $\beta _{r}$, $\gamma_{r}$, and the product 
$\gamma _{0}\beta _{0}$ are equal or greater than zero, one has $\left(
\varepsilon +m\right)\left( \varepsilon -m\right)>0$, meaning that either $%
\varepsilon >m$ or $\varepsilon <-m$, a feature characteristic of harmonic
oscillator states for positive energy (particle) or for negative energy
(anti-particle) states. Furthermore, since the products $\pm \gamma _{s}(
\varepsilon \pm m)$ and $\pm \alpha_{s}( \varepsilon \pm m)$ under the
square roots must be positive for arbitrary high values of $|\varepsilon|$,
both $\alpha_s$ and $\gamma_s$ must be positive for $V_s=V_v$ and negative
for $V_s=-V_v$ for positive energy states, that is, $V_v$ is always positive
for these states. The reverse is true for negative energy states.

Squaring Eq. (\ref{Na}) successively results into a nonequivalent polynomial
equation of degree $8$. Solutions of this algebraic equation that are not
solutions of the original equation can be removed by backward substitution.
A quartic algebraic equation is obtained when $\alpha _{s}=0$ (the case of a
nonminimal vector\ Cornell potential plus an equal-magnitude mixing of
vector and scalar harmonic oscillators). For $\alpha _{s}=\gamma _{r}\left(
2\beta _{r}+1+2k\right) -2\gamma _{0}\beta _{0}=\gamma _{r}^{2}-\gamma
_{0}^{2}=0$ (which includes the case of a nonminimal vector\ Coulomb
potential and the case of a space component of a nonminimal vector\ Coulomb
potential plus a time component of a nonminimal vector\ linear potential,
both plus an equal-magnitude mixing of vector and scalar harmonic
oscillators) one obtains a cubic algebraic equation. However, (\ref{Na}) can
be written as a quadratic algebraic equation rendering two branches of
solutions symmetrical about $\varepsilon =0$ in the case of a pure
nonminimal vector Cornell potential ($\alpha _{s}=\gamma _{s}=0\,,|\gamma
_{r}|\neq |\gamma _{0}|$):%
\begin{equation}
\varepsilon =\pm \sqrt{m^{2}+\gamma _{r}\left( 2\beta _{r}+1+2k\right)
-2\beta _{0}\gamma _{0}+2\sqrt{\gamma _{r}^{2}-\gamma _{0}^{2}}\left(
2n+1+S\right) }\,,
\end{equation}%
where $S=\sqrt{\left( l+k-1/2\right) ^{2}+\beta _{r}\left( \beta
_{r}-1+2k\right) -\beta _{0}^{2}}$. One may also note that the (positive
energy) $D$-dimensional relativistic harmonic oscillator is obtained when $%
\gamma _{r}=\gamma _{0}=\beta _{r}=\beta _{0}=\alpha _{s}=0$ and $\gamma
_{s}=\pm (1/2)\,m\Omega ^{2}$ for $V_{v}=\pm V_{s}$, where $\Omega $ is the
harmonic oscillator frequency. In this case $B=C=0$ and one obtains $M\omega
=\sqrt{m\Omega ^{2}\left( \varepsilon \pm m\right) }$ and $\varepsilon
^{2}-m^{2}=2(2n+1+\left\vert l+k-\frac{1}{2}\right\vert )\sqrt{m\Omega
^{2}\left( \varepsilon \pm m\right) }$. The Klein-Gordon $3D$ oscillator is
obtained of course as a particular case with $k=1$. At this point, it is
worth remarking that, when there are only scalar and vector potentials, the
conditions $V_{v}=\pm V_{s}$ correspond to having spin(pseudospin) symmetry
conditions in the Dirac equation with $D=3$ (a recent review of this subject
is given in \cite{liang2015hidden}) and that, under those conditions,
the energy spectrum of the Dirac and the Klein-Gordon equations is the same 
\cite{alberto2007spin}.

\subsubsection{Vector-scalar SHO plus nonminimal vector Coulomb (space
component) and harmonic oscillator (time component) potentials}

Another example is given by%
\begin{equation}
V_{r}=\frac{\beta _{r}}{r},\quad V_{0}=\gamma _{0}r^{2},\quad V_{s}=\frac{%
\alpha _{s}}{r^{2}}+\gamma _{s}r^{2},\quad V_{v}=\pm \frac{\alpha _{s}}{r^{2}%
}+\gamma _{v}r^{2},
\end{equation}%
with $\gamma _{s}^{2}=\gamma _{v}^{2}+\gamma _{0}^{2}$ and 
\begin{eqnarray}
M\omega &=&\sqrt{2\left( \varepsilon \gamma _{v}+m\gamma _{s}\right) } 
\notag \\
2MB &=&\beta _{r}\left( \beta _{r}-1+2k\right) \pm 2\alpha _{s}\left(
\varepsilon \pm m\right) \\
2MC &=&2\alpha _{s}\left( \gamma _{s}\mp \gamma _{v}\right) .  \notag
\end{eqnarray}%
In this case, the quantization condition%
\begin{equation}
\left( \varepsilon +m\right) \left( \varepsilon -m\right) -2\alpha
_{s}\left( \gamma _{s}\mp \gamma _{v}\right) =2\left( 2n+1+S\right) \sqrt{%
2\left( \varepsilon \gamma _{v}+m\gamma _{s}\right) }
\end{equation}%
where $S=\sqrt{\left( l+k-1/2\right) ^{2}+\beta _{r}\left( \beta
_{r}-1+2k\right) \pm 2\alpha _{s}\left( \varepsilon \pm m\right) }$, is
again convertible to an algebraic equation of degree $8$. A quartic
algebraic equation is obtained when $\alpha _{s}=0$, and a quadratic
algebraic equation when $\alpha _{s}=\gamma _{v}=0$:%
\begin{equation}
\varepsilon =\pm \sqrt{m^{2}+2\sqrt{2m\gamma _{s}}\left( 2n+1+S\right) },
\end{equation}%
with $m\neq 0$ and $\gamma _{s}>0$. In this case, $|\gamma _{0}|=\gamma _{s}$
and the spectrum is indifferent to the sign of $\gamma _{0}$.

\subsection{The effective singular Coulomb potential}

Comparison of the bound states of Eq. (\ref{Eq}) with those ones from the
generalized Morse potential with the pair $(\delta ,\Lambda )=(-1,1)$ is
done by choosing $V_{1}=\alpha ^{2}r_{0}A$ and $V_{2}=-\alpha
^{2}r_{0}^{2}\left( \widetilde{\varepsilon }-C\right) $, with $A<0$ and $%
\widetilde{\varepsilon }<C$. Now, 
\begin{equation}
\xi =2\sqrt{2M\left( C-\widetilde{\varepsilon }\right) }\,r
\end{equation}%
and (\ref{cond}) implies $\widetilde{\varepsilon }>C-MA^{2}/[2\left(
n+1/2\right) ^{2}]$. Using (\ref{ENE}) and (\ref{etil}) one can write

\begin{eqnarray}
\widetilde{\varepsilon } &=&C-\frac{MA^{2}}{2\zeta ^{2}}  \notag \\
&& \\
u\left( r\right) &=&Nr^{1/2+S}e^{-M|A|r/\zeta }L_{n}^{\left( 2S\right)
}\left( \frac{2M|A|r}{\zeta }\right) ,  \notag
\end{eqnarray}%
with%
\begin{equation}
\zeta =n+1/2+S.
\end{equation}
Again, we see that there is no upper limit on the value of $n$. The spectrum
is bound from below and above, since, when $n\to\infty$, $\widetilde{%
\varepsilon }\to C$, corresponding to energies $\varepsilon=\pm\sqrt{m^2+2MC}
$, in which case there are no bound states, and of course one must have $%
2MC>-m^2$. One has in general 
\begin{equation}  \label{Coul_energ_spectr}
2MC-\left(\frac{2MA}{2S+1}\right)^2+m^2\leq \varepsilon^2 < 2MC+m^2\ .
\end{equation}
This is the characteristic behaviour of Coulomb-like spectra.

\subsubsection{Vector-scalar Coulomb plus nonminimal vector Cornell
potentials}

An example of this class of solutions can be reached by choosing%
\begin{equation}
V_{r}=\frac{\beta _{r}}{r}+\gamma _{r}r,\quad V_{0}=\frac{\beta _{0}}{r}\pm
\gamma _{r}r,\quad V_{s}=\frac{\beta _{s}}{r},\quad V_{v}=\frac{\beta _{v}}{r%
}.
\end{equation}%
This class generalizes the results found in \cite{zhang2009dynamical} ($V_{r}=V_{0}=0$ and $%
V_{v}=V_{s}=\beta _{s}/r$ in two dimensions), \cite{dong2011wave,Hassanabadi2011,saad2008klein} ($V_{r}=V_{0}=0$ in $D$ dimensions).

There results%
\begin{eqnarray}
2MA &=&2\left( m\beta _{s}+\varepsilon \beta _{v}\right)  \notag \\
2MB &=&\beta _{s}^{2}-\beta _{v}^{2}+\beta _{r}\left( \beta _{r}-1+2k\right)
-\beta _{0}^{2} \\
2MC &=&\gamma _{r}\left( 2\beta _{r}+1+2k\mp 2\beta _{0}\right) ,  \notag
\end{eqnarray}%
so that%
\begin{equation}
\varepsilon =\frac{-m\beta _{v}\beta _{s}\pm \zeta \sqrt{\left( \beta
_{v}^{2}+\zeta ^{2}\right) \left[ m^{2}+\gamma _{r}\left( 2\beta
_{r}+1+2k\mp 2\beta _{0}\right) \right] -m^{2}\beta _{s}^{2}}}{\beta
_{v}^{2}+\zeta ^{2}},
\end{equation}%
necessarily with $\varepsilon ^{2}<m^{2}+\gamma _{r}\left( 2\beta
_{r}+1+2k\mp 2\beta _{0}\right) .$

\subsubsection{Vector-scalar Coulomb plus nonminimal vector shifted Coulomb
potentials}

Another example, generalizing the results found in \cite{zhang2009dynamical} ($V_{r}=V_{0}=0
$ and $V_{v}=V_{s}=\beta _{s}/r$ in two dimensions), \cite{dong2011wave,Hassanabadi2011,saad2008klein} ($V_{r}=V_{0}=0$ in $D$ dimensions), is%
\begin{equation}
V_{r}=\frac{\beta _{r}}{r}+\gamma _{r},\quad V_{0}=\frac{\beta _{0}}{r}%
+\gamma _{0},\quad V_{s}=\frac{\beta _{s}}{r},\quad V_{v}=\frac{\beta _{v}}{r%
},
\end{equation}%
resulting in%
\begin{eqnarray}
2MA &=&2\left[ m\beta _{s}+\varepsilon \beta _{v}+\gamma _{r}\left( \beta
_{r}+k\right) -\gamma _{0}\beta _{0}\right]   \notag \\
2MB &=&\beta _{r}\left( \beta _{r}-1+2k\right) -\beta _{0}^{2}+\beta
_{s}^{2}-\beta _{v}^{2} \\
2MC &=&\gamma _{r}^{2}-\gamma _{0}^{2},  \notag
\end{eqnarray}%
so that, defining%
\begin{equation}
\tau =m\beta _{s}+\gamma _{r}\left( \beta _{r}+k\right) -\gamma _{0}\beta
_{0}
\end{equation}%
one finds%
\begin{equation}
\varepsilon =\frac{-\beta _{v}\tau \pm \zeta \sqrt{\left( \beta
_{v}^{2}+\zeta ^{2}\right) \left( m^{2}+\gamma _{r}^{2}-\gamma
_{0}^{2}\right) -\tau ^{2}}}{\beta _{v}^{2}+\zeta ^{2}},
\end{equation}%
with $\varepsilon ^{2}<m^{2}+\gamma _{r}^{2}-\gamma _{0}^{2}$. The pure
nonminimal vector\ shifted Coulomb potential holds the spectrum%
\begin{equation}
\varepsilon =\pm \sqrt{m^{2}+\gamma _{r}^{2}-\gamma _{0}^{2}-\left[ \frac{%
\gamma _{r}\left( \beta _{r}+k\right) -\gamma _{0}\beta _{0}}{\zeta }\right]
^{2}}.
\end{equation}%
Curiously, there can be solutions even if $\beta _{r}=\beta _{0}=0$ with $%
\gamma _{r}<0$, i.e., one has bound solutions only with constant nonminimal
vector potentials. Note that in this example, \textquotedblleft $\pm $%
\textquotedblright\ denotes two different solutions regarding the same
potentials parameters.

\subsubsection{Vector-scalar SCP plus nonminimal vector Coulomb potentials}

Now consider the case of a nonminimal vector\ Coulomb potential plus an
equal-magnitude mixing of vector and scalar SCP:%
\begin{equation}
V_{r}=\frac{\beta _{r}}{r},\quad V_{0}=\frac{\beta _{0}}{r},\quad V_{s}=%
\frac{\alpha _{s}}{r^{2}}+\frac{\beta _{s}}{r},\quad V_{v}=\pm V_{s}.
\end{equation}%
Chiefly due to the presence of the nonminimal vector\ Coulomb potential,
this last example generalizes those ones found in Refs. \cite{saad2008klein} (SCP in $D
$ dimensions), \cite{kocak2007bound} (Kratzer potential for $S$-waves in three
dimensions, as a matter of fact the SCP), \cite{berkdemir2007relativistic} (Kratzer potential with 
$V_{v}=V_{s}$ in three dimensions), and \cite{wen2003bound} (SCP with $V_{v}=V_{s}$
in three dimensions). One finds 
\begin{eqnarray}
2MA &=&\pm 2\beta _{s}\left( \varepsilon \pm m\right)   \notag \\
2MB &=&\beta _{r}\left( \beta _{r}-1+2k\right) -\beta _{0}^{2}\pm 2\alpha
_{s}\left( \varepsilon \pm m\right)  \\
2MC &=&0,  \notag
\end{eqnarray}%
in such a way that for $m\neq 0$ one finds the irrational equation in $%
\varepsilon $%
\begin{flalign}
\left( \varepsilon +m\right)& \left( \varepsilon -m\right)
= -\left[ \frac{%
\beta _{s}\left( \varepsilon \pm m\right) }{\zeta }\right] ^{2} \notag \\
  & = -\left[ \frac{%
\beta _{s}\left( \varepsilon \pm m\right) }
{n+1/2+\sqrt{\left( l+k -1/2\right) ^{2}+
\beta _{r}\left( \beta _{r}-1+2k \right) -\beta _{0}^{2}\pm
2\alpha _{s}\left( \varepsilon \pm m\right)} }\right] ^{2},
\end{flalign}
necessarily with $|\varepsilon |<m$ and $\beta _{s}<0$.

\subsubsection{\protect\bigskip Vector-scalar SCP plus nonminimal shifted
Coulomb and inverse square root potentials}

An additional example containing SCP potentials is given by 
\begin{equation}
V_{r}=\frac{\beta _{r}}{r}+\gamma _{r},~V_{0}=\frac{\gamma _{0}}{\sqrt{r}}%
,~V_{s}=\frac{\alpha _{s}}{r^{2}}+\frac{\beta _{s}}{r},~V_{v}=\pm V_{s},
\end{equation}%
resulting in%
\begin{eqnarray}
2MA &=&2\gamma _{r}\left( \beta _{r}+k\right) -\gamma _{0}^{2}\pm 2\beta
_{s}\left( \varepsilon \pm m\right)  \notag \\
2MB &=&\beta _{r}\left( \beta _{r}-1+2k\right) \pm 2\alpha _{s}\left(
\varepsilon \pm m\right) \\
2MC &=&\gamma _{r}^{2},  \notag
\end{eqnarray}%
and giving, after properly squaring the original equation, an algebraic
equation of degree 6 in $\varepsilon $. For $\alpha _{s}=0$, though, one
obtains a quadratic algebraic equation with the solutions 
\begin{equation}
\varepsilon =\frac{-\beta _{v}\tau \pm \sqrt{\beta _{v}^{2}\tau ^{2}+\left(
\beta _{v}^{2}+\zeta ^{2}\right) \left[ 4\zeta ^{2}\left( m^{2}+\gamma
_{r}^{2}\right) -\tau ^{2}\right] }}{2\left( \beta _{v}^{2}+\zeta
^{2}\right) },
\end{equation}%
where%
\begin{equation}
\tau =2\left[ \gamma _{r}\left( \beta _{r}+k\right) -\gamma _{0}^{2}/2+\beta
_{s}m\right]
\end{equation}%
In the case of a pure nonminimal vector potential one finds%
\begin{equation}
\varepsilon =\pm \sqrt{m^{2}+\gamma _{r}^{2}-\left[ \tau /\left( 2\zeta
\right) \right] ^{2}}.
\end{equation}%
%
%
%
%
%
%

\subsubsection{A special case}

The very special case%
\begin{equation}
V_{r}=\frac{\beta _{r}}{r},\quad V_{0}=\frac{\beta _{0}}{r},\quad V_{s}=%
\frac{\beta _{s}}{r},\quad V_{v}=\pm V_{s}
\end{equation}%
consists in particular cases of the first three preceding cases, yielding a
spectrum for $m\neq 0$ given by%
\begin{equation}
\varepsilon =\pm m\frac{1-\left( \beta _{s}/\zeta \right) ^{2}}{1+\left(
\beta _{s}/\zeta \right) ^{2}}.
\end{equation}%
This spectrum is formally identical to the relativistic Coulomb potential
for Klein-Gordon and Dirac equations in the condition of spin or pseudospin
symmetry mentioned before ($V_{v}=\pm V_{s}$). The reason is that the
nonminimal Coulomb potentials enter only in the expression of $\zeta $,
which is given by 
\begin{equation}
\zeta =n+1/2+\sqrt{\left( l+k-1/2\right) ^{2}+\beta _{r}\left( \beta
_{r}-1+2k\right) -\beta _{0}^{2}}\ .  \label{eta}
\end{equation}%
When there are only scalar and vector potentials $\zeta =n+1/2+|l+k-1/2|=N$
and one recovers the expressions for the energy of a $D$-dimensional
relativistic Coulomb spin-0 system in spin or pseudospin symmetry conditions
(if $D$ is odd, $N$ is an integer, and for $D=3$ it is the principal quantum
number of the non-relativistic hydrogen atom). One may note that, contrary
to general case of some examples discussed before, here the choice $%
V_{s}=V_{v}$ (spin condition) implies that there is only one (positive)
energy bound solution and the other choice $V_{s}=-V_{v}$ (pseudospin
condition) means that there is only a negative energy bound solution. As
might be expected, this is exactly what happens in the corresponding
(3-dimensional) Dirac equation \cite{dde2012spine2012spin}.

\section{Concluding remarks}

Based on Ref. \cite{Nogueira2016a}, we have described a straightforward and efficient
procedure for finding a large class of new solutions of the $D$-dimensional
Klein-Gordon equation with radial scalar, vector, and nonminimal coupling
potentials, whose wave functions are all expressed in terms of generalized
Laguerre polynomials and whose energy eigenvalues obey analytical equations,
either polynomial or irrational which can be cast as polynomial. These
include harmonic oscillator-type and Coulomb-type potentials and their
extensions. Although the solutions for those systems could be found by
standard methods, this procedure, based on the mapping from the
one-dimensional generalized Morse potential via a Langer transformation to
the $D$-dimensional radial Klein-Gordon equation, provide an easier and
powerful way to find the solutions of a very general class of potentials
which otherwise one might not know that would have analytical solutions in
the first place. We were able to reproduce well-known particular cases of
relativistic harmonic oscillator and Coulomb spin-0 systems, especially when
the scalar and vector potentials have the same magnitude, but there are a
wealth of other particular cases with physical interest that are left for
further study.
\cite{zhang2009dynamical}

\section*{Acknowledgement}

This work was supported in part by means of funds provided by CAPES and CNPq
(grants 455719/2014-4, 304105/2014-7 and 304743/2015-1). PA would like to
thank the Universidade Estadual Paulista, Guaratinguet\'a Campus, for
supporting his stays in its Physics and Chemistry Department. 
\section*{References}

\bibliography{v4MORSEkg}

\end{document}